\documentclass[aps,prl,twocolumn]{revtex4-1}
\usepackage{amsmath}
\usepackage{amsfonts}
\usepackage{graphicx}
\usepackage{color}

\begin{document}

\title{Majorana bound states in two-channel time-reversal-symmetric nanowire systems}
\author{Erikas Gaidamauskas, Jens Paaske, and Karsten Flensberg}
\affiliation{Center for Quantum Devices, Niels Bohr Institute,
University of Copenhagen, Universitetsparken 5, DK-2100 Copenhagen,
Denmark}

\date{\today}

\begin{abstract}
 We consider time-reversal-symmetric two-channel semiconducting quantum wires proximity coupled to an $s$-wave superconductor. We analyze the requirements for a nontrivial topological phase and find that necessary conditions are 1) the determinant of the pairing matrix in channel space must be negative, 2) inversion symmetry must be broken, and 3) the two channels must have different spin-orbit couplings. The first condition can be implemented in semiconducting nanowire systems where interactions suppress intra-channel pairing, while the inversion symmetry can be broken by tuning the chemical potentials of the channels. For the case of collinear spin-orbit directions, we find a general expression for the topological invariant by block diagonalization into two blocks with chiral symmetry only. By projection to the low-energy sector, we solve for the zero modes explicitly and study the details of the gap closing, which in the general case happens at finite momenta.
\end{abstract}

\pacs{71.10.Pm, 74.45.+c, 74.78.Na}

\maketitle

Majorana fermionic bound states (MBS) are theoretically predicted to exist at the boundaries of topological superconducting states\cite{Kitaev2001} and to have non-Abelian exchange statistics\cite{Ivanov2001}. They are, therefore, promising proposals for realizations of elements of topological quantum computation\cite{Nayak2008} and currently there is an extensive search for candidate systems. Promising suggestions are hybrid condensed matter systems with $s$-wave superconductors proximity coupled to materials with strong spin-orbit coupling \cite{AliceaReview,BeenakkerReview,LeijnseReview,HasanKaneReview}. Recent theoretical proposals \cite{Fu2008,Lutchyn2010,Oreg2010} for 1D and quasi-1D \cite{Potter2010,Lutchyn2011} topological superconducting systems and first experimental results\cite{Mourik2012,Das2012,Churchill2013} have received wide interest. Interestingly, the non-Abelian nature of the Majorana bound states can be explored also in 1D systems in a wire-network geometry \cite{Alicea2011,Halperin2012}.

All of the above refers to superconducting systems in the topological symmetry class D\cite{HasanKaneReview}, where breaking of time-reversal symmetry (TRS) leads to a single localized MBS. With additional symmetry (BDI class) multiple nondisordered protected MBS are also possible in multichannel systems\cite{Fulga2011,Stanescu2011}. Recent papers have considered the possibility of realizing 1D topological superconductor systems with time-reversal symmetry (class DIII), supporting Majorana Kramers doublets in hybrid structures based either on superconductors with $d_{x^2-y^2}$-wave\cite{Wong2012} or $s_{\pm}$-wave\cite{Zhang2013} pairing, noncentrosymmetric superconductors\cite{Nakosai2013}, bilayer 2D superconductors with spin-orbit coupling\cite{Nakosai2012}, or on 1D two-band models with conventional $s$-wave supercoductor\cite{Deng2012,Keselman2013} under the assumption of a $\pi$ phase difference between the pairing potentials in the two bands, mimicking the $s_\pm$ pairing considered in Ref.~\onlinecite{Zhang2013}.
It is interesting to note that even though two local MBS together form a usual fermion, the exchange of two Kramers pairs of MBS also constitutes a non-Abelian operation\cite{Liu2013}. Moreover, just as for single MBS, the Kramers MBS can be detected either by tunneling spectroscopy or via unusual current-phase relations in a Josephson junction to an ordinary $s$-wave superconductor\cite{Chung2012}.
\begin{figure}
\centering
\includegraphics[scale=0.55]{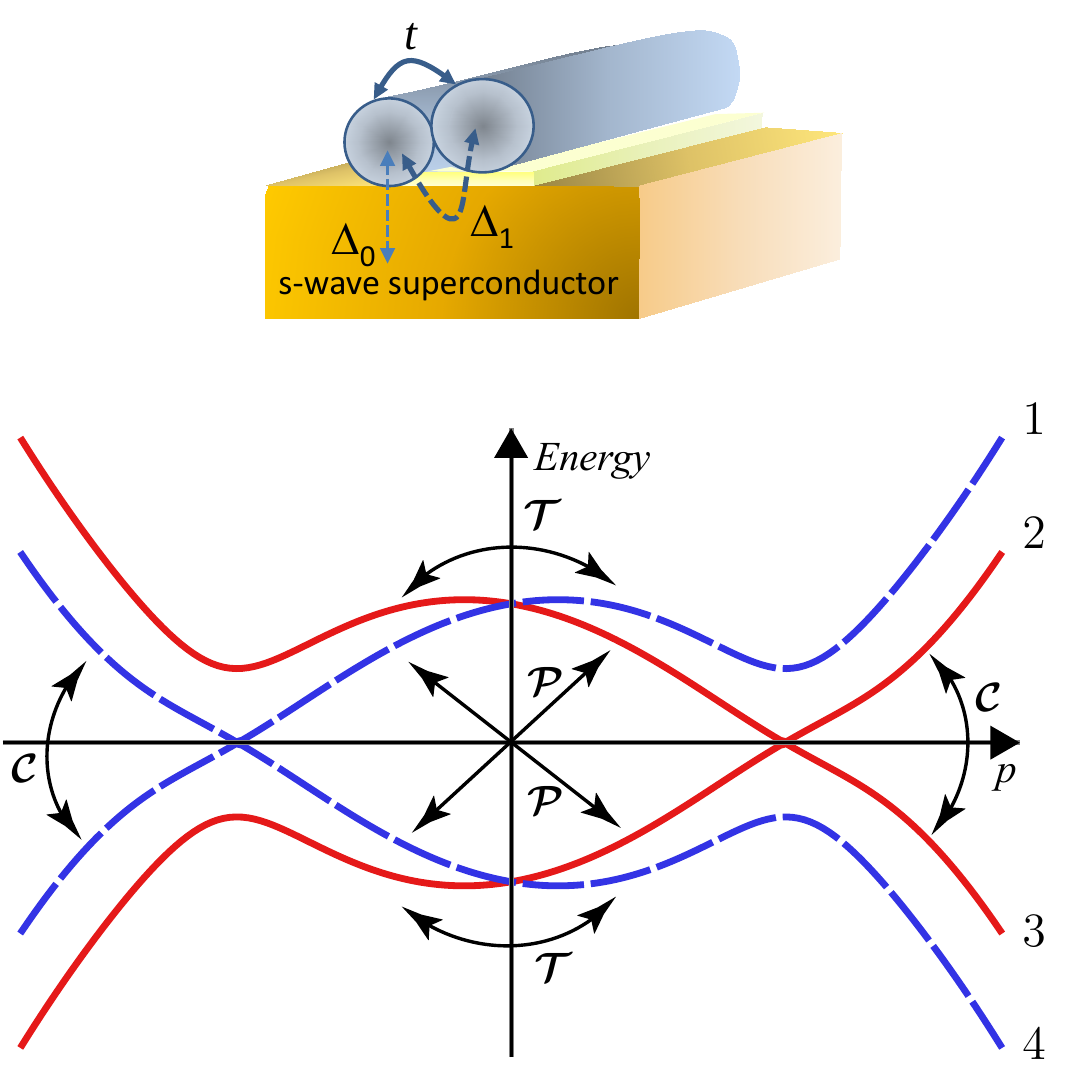}
\caption{(color online). {Top:}. Sketch of the geometry of the two-channel (or two-wire) superconducting system. Bottom: General structure of the low-energy bands at the gapless transition point. Because the blocks in the block-diagonal Hamiltonian have chiral symmetry, the bands that cross at zero energy are related by
$\mathcal{C}$, which means $E_2(p)=-E_3(p)$, while the bands that cross at $p=0$ are related by TRS, $\mathcal{T}:\, E_1(p)=E_2(-p)$. Finally, particle-hole symmetry implies $\mathcal{P}:\, E_2(p)=-E_4(-p)$. This plot is generated using the model in Eq.~\eqref{eq1}, which assumes collinear spin-orbit coupling. However, it should be emphasized that the general structure is preserved even without this assumption.
\label{fig:geo}}
\end{figure}

In this Letter, we investigate a model with two channels coupled to an $s$-wave superconductor, see Fig.~\ref{fig:geo}. The two channels {\it wires}) could be either two transverse modes in a single nanowire or separate nanowires as illustrated in Fig.~1.
We demonstrate that {\it interwire pairing} can give rise to a topologically nontrivial phase with Kramers MBS at the ends. This relies only on having different spin-orbit coupling in  the two wires, and on the square of the interwire pairing being larger than the product of the intrawires pairings. In a conventional (i.e., constant phase) $s$-wave superconductor, the latter condition cannot be achieved without interactions [21]. However, intrawire pairing can be significantly suppressed by repulsive intrawire interactions\cite{Gangadharaiah2011,Stoudenmire2011,Recher2002}, thus enhancing the role of the interwire pairing.

Our analysis is carried out under the simplifying assumption of collinear spin-orbit coupling axis in the two wires. However, the structure of the low lying energy bands (shown in Fig.~\ref{fig:geo}) relies entirely on the TRS and is, therefore, preserved even when relaxing this assumption. In the case of collinearity, the Hamiltonian can be block diagonalized into two blocks related by time reversal $\mathcal{T}$ and particle-hole $\mathcal{P}$ transformation. Each individual block has merely chiral symmetry $\mathcal{C}$ and is simple enough that we can give an analytical expression for the corresponding (class AIII) topological invariant, which changes with the gap closings.

The Bogoliubov--de Gennes Hamiltonian for the TRS two-channel nanowire system is
\begin{equation}
\label{eq1}
H_{BdG} = \frac{1}{2} \int_{- \infty}^{\infty} \Psi^{+} (x)\mathcal{H} \Psi (x) dx,
\end{equation}
where the first-quantization Hamiltonian is ($\hbar=1$)
\begin{align}
\label{eq2}
\mathcal{H} =& \left(\frac{p^2}{2m} - \mu + V \lambda_{z} +t\lambda_x+(\alpha+\gamma\lambda_x+\beta\lambda_z) p \sigma_{z}\right) \tau_{z} \notag\\
&+ (\Delta_0 + \Delta_3 \lambda_{z}+\Delta_1 \lambda_{x})\tau_{x},
\end{align}
where $\mu$ is the chemical potential, $V$ is the difference in electrical potentials, $\alpha$ and $\beta$ are the symmetric and antisymmetric parts of the spin-orbit coupling coefficients, {$\gamma$ is the interwire spin-orbit coupling, and $\Delta_0\pm \Delta_3$ and $\Delta_1$ are the intrawire and interwire pairing potentials, respectively. Pauli matrices $\sigma,\lambda$, and $\tau$ act on the three two-dimensional spaces: spin, wire index, and electron-hole, respectively. In writing Eq.~\eqref{eq2}, we have used the conventional Nambu basis: $\Psi(x)=(\Psi_{\uparrow}(x),\Psi_{\downarrow}(x),\Psi_{\downarrow}^\dagger(x),-\Psi_{\uparrow}^\dagger(x))$.

The Hamiltonian \eqref{eq2} belongs to the topological symmetry class DIII with both antiunitary particle-hole and time-reversal symmetries, and hence unitary chiral symmetry \cite{sym,HasanKaneReview}. In our basis, $ \mathcal{T} = i \sigma_{y} \mathcal{K}$, $\mathcal{P} = \sigma_{y} \tau_{y} \mathcal{K}$, and $\mathcal{C}=i \mathcal{T}\mathcal{P}$, where $\mathcal{K}$ is complex conjugation. Because the Hamiltonian is block diagonal in spin space, we can write it as
\begin{equation}\label{Hblock}
  \mathcal{H}=\left(
                \begin{array}{cc}
                  \mathcal{H}_{p,\uparrow} & 0 \\
                  0 & \mathcal{H}_{p,\downarrow} \\
                \end{array}
              \right),\quad
\mathcal{H}_{p,\sigma}=\left(
                \begin{array}{cc}
                  \mathcal{H}_{p,\sigma}^0 & \Delta \\
                  \Delta & -\mathcal{H}_{p,\sigma}^0 \\
                \end{array}
              \right),
\end{equation}
where $ \mathcal{H}_{p,\sigma}^0$ and $\Delta$ are 2$\times$2 matrices in wire-index space. The two blocks in Eq.~\eqref{Hblock} are related by time-reversal and particle-hole symmetry, $\mathcal{T} \mathcal{H}_{p,\uparrow} \mathcal{T}^{-1}=\mathcal{H}_{-p,\downarrow}$ and  $\mathcal{P} \mathcal{H}_{p,\uparrow} \mathcal{P}^{-1}=-\mathcal{H}_{-p,\downarrow}$, which means that each block only has chiral symmetry $\mathcal{C} \mathcal{H}_{p,\sigma} \mathcal{C}^{-1}=-\mathcal{H}_{p,\sigma}$. Considered separately, the Hamiltonian in each block belongs to symmetry class AIII\cite{HasanKaneReview}. The gap of the spectrum of $\mathcal{H}$ vanishes for certain parameters, indicating a potential topological transition. The gap closing happens at finite momenta, which distinguishes this system from the above-mentioned $s_\pm$-wave pairing models \cite{keselmannote}. This is illustrated in Fig.~\eqref{fig:geo} which shows the generic situation for the low energy bands at the point where the gap closes.
\begin{figure}
\centering
\includegraphics[width=.5\textwidth]{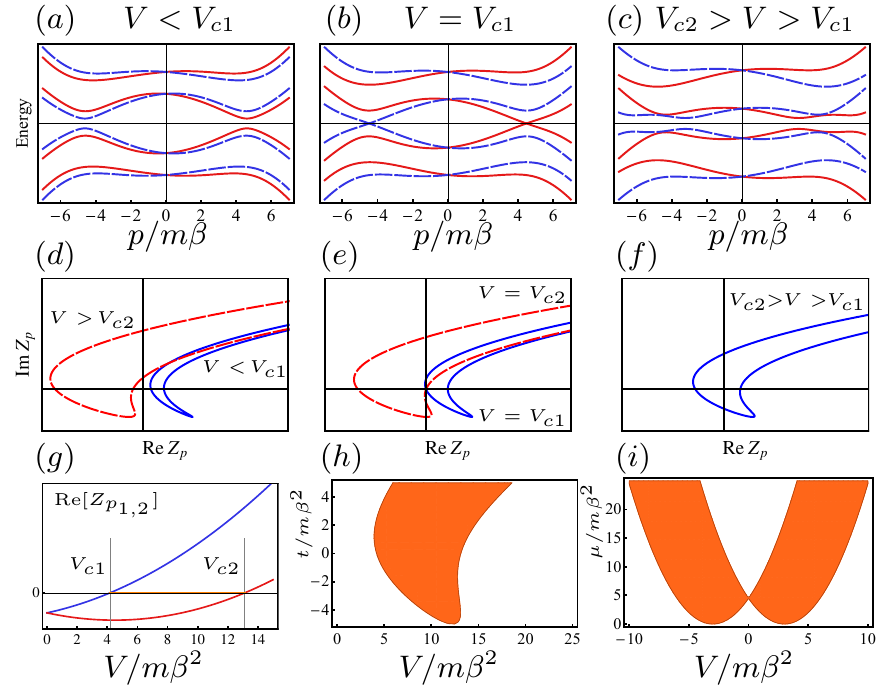}
\caption{(color online). {Top:} (a)-(c) Spectrum of the two-wire model for three values of the potential difference, $V$. In (a), the system is in the trivial state, $V<V_{c1}$, (b) is at the transition point, $V=V_{c1}$, while (c) shows the gapped spectrum in the topological phase. (d)-(f) The contour followed by the complex determinants $Z_{p}$ from negative to positive values of $p$. In (d) the system is in the trivial state, which means $V$ smaller than the lowest critical value $V_{c1}$ or larger than the
largest critical value $V_{c2}$. In (e) the contours go through the origin, which means that the gap closes at both $V=V_{c1}$ and $V=V_{c2}$, while in (f) the contour encircles the origin, signifying the topological state. The constant parameters in the plots are $\alpha=\gamma=0,\Delta_1=10m\beta^2,\Delta_0=5m\beta^2,t=0$, and $\mu=10 m\beta^2$, which gives $V_{c1}=4.19 m\beta^2$ and $V_{c2}=13.13 m\beta^2$. (g)-(i) Examples of the topological phase space. In (g) the real part of the $Z_p$ determinant at $p=p_1$ and $p=p_2$ is shown as a function of $V$. For the topological criterion to be fulfilled the product of the two function must be negative, which occurs for $V$ between $V_{c1}$ and $V_{c2}$. (h) Phase diagram when varying the potential $V$ and the interwire tunneling $t$ (same constant parameters as above), and (i) a phase diagram in the $V$-$\mu$ space (same as above except $\Delta_1=5m\beta^2$ and $\Delta_0=4 m\beta^2$).
\label{fig:phases}}
\end{figure}

To establish that the closing and reopening of the gap is associated with a topological transition, a topological invariant is required. Since $\mathcal{H}_{p,\sigma}$ (in AIII) lacks particle-hole symmetry, a Pfaffian cannot be defined as for class D systems\cite{Kitaev2001}. Nevertheless, we can still extract information about the sign of the gap from the square root of the determinant of the Hamiltonian. Transforming $\mathcal{H}_{p,\sigma}$ to
\begin{equation}\label{Hblocksigma}
U\mathcal{H}_{p,\sigma}U^\dagger=\left(
                \begin{array}{cc}
                  0 & \Delta-i\mathcal{H}_{p,\sigma}^0 \\
                  \Delta+i\mathcal{H}_{p,\sigma}^0 &  0\\
                \end{array}
              \right),
\end{equation}
using $U=\exp(i\tau_x\pi/4)$, the determinant reads
\begin{equation}\label{det}
 \det(\mathcal{H})=|\det(\Delta+i\mathcal{H}_{p,\uparrow}^0)|^2|
 \det(\Delta+i\mathcal{H}_{p,\downarrow}^0)|^2,
\end{equation}
suggesting that the sign of the gap is encoded in the function $Z_p=\det(\Delta+i\mathcal{H}_{p,\uparrow}^0)
=\det(\Delta+i\mathcal{H}_{-p,\downarrow}^0)$. In fact, the winding number of $z_p=Z_p/|Z_p|=\exp(i\theta_p)$ defined as
\begin{equation}\label{W}
  W=\frac{1}{2\pi i}\int_{p=-\infty}^{p=\infty} \frac{dz}{z}=\frac{1}{2\pi}\int_{-\infty}^{\infty}dp\,  \frac{d \theta_p}{dp},
\end{equation}
takes only integer values since $z_{p=\infty}=z_{p=-\infty}$, and the topological invariant associated with the $\mathbb{Z}_2$ classification of the full class DIII Hamiltonian is given by $Q=(-1)^W$, similarly to the analysis by Tewari and Sau for BDI symmetry class models\cite{Tewari2012}. Nontrivial values of the winding number (topological invariant) is always related to the changes in the topology of the gapped system. In our model it corresponds to the number of Majorana bound states at each end of the nanowire.

Since, however, the winding number in Eq.~\eqref{W} is not well suited for analytical evaluation, we shall instead determine the condition for a topological state directly from the determinant $Z_p$. This is done by first identifying the $p$ values at which $\mathrm{Im} Z_p=0$, giving two solutions:
\begin{align}\label{ppm}
  &p_{1(2)}=-m\alpha+\frac{m\gamma\Delta_1}{\Delta_0}+\frac{m\beta\Delta_3}{\Delta_0}\pm p_0,\quad \\
  &p_0^2={2m\left(\mu+\frac{t\Delta_1}{\Delta_0}+
  \frac{V\Delta_3}{\Delta_0}\right)
  +m^2\left(\alpha+\gamma\frac{\Delta_1}{\Delta_0}-\beta\frac{\Delta_3}{\Delta_0}\right)^2}.\notag
\end{align}
Therefore, when $p$ runs from $-\infty$ to $+\infty$ the complex number $Z_p$ crosses the real axis exactly two times and encloses the origin if and only if
\begin{equation}\label{Qdef}
  Q=\mathrm{sign}\left[Z_{p_1}Z_{p_2}\right]=-1.
\end{equation}

We can now draw some general conclusions. Firstly, it is straightforward (see Appendix) to show that the eigenvalues of the pairing matrix $\Delta$ must have different signs in order to have $Q=-1$. In other words, one must have $\det\Delta=\Delta_0^2-\Delta_3^2-\Delta_1^2<0$. Secondly, if we define an inversion symmetry by  $\mathcal{I}=\lambda_x$, the Hamiltonian is inversion symmetric if $\mathcal{I}H(p)\mathcal{I}=H(-p)$. Setting the terms that break inversion symmetry to zero, i.e., $V=\alpha=\gamma=\Delta_3=0$, it can be seen that $Q=1$. Therefore, inversion symmetry must be broken in order to have a topologically nontrivial phase. Finally, it follows that $Q=1$ if $\gamma=\beta=0$, which means that the spin-orbit matrix $\alpha+\gamma\lambda_x+\beta\lambda_z$ must have two different eigenvalues.

The full expression for the topological quantum number $Q$ can be found algebraically, but is in general rather involved. Therefore, we present some special cases in the following. First, we write the result for the case $\Delta_3=\gamma=0$:
\begin{align}\label{Qz0}
Q_{\Delta_3=0}=&\mathrm{sign}\left[A^2-B^2\right],\\
A=&\Delta_0^2(V^2+\delta^2-2m\alpha\beta V+\beta^2(p_0^2+m^2\alpha^2))+t^2\delta^2,\notag\\
B=&2\Delta_0^2\beta p_0(m\alpha\beta-V),\notag
\end{align}
where $\delta^2=\Delta_0^2-\Delta_1^2$. From this it is evident that $\beta\neq 0$ is a necessary condition for a nontrivial phase, in agreement with the above general conclusion. If we further take $\alpha=t=0$, the condition becomes
\begin{equation}\label{KK}
K_{-} < \Delta_1 < K_{+},
\end{equation}
with $K_{\pm} = \sqrt{(V\pm \sqrt{2\beta^{2}\mu m})^2 +\Delta_0^{2}}$. Clearly, this expression requires $\Delta_1>\Delta_0$, which (as discussed in the introduction) could be realized due to repulsive interactions. Below, we look at the more general case of different intrawire pairings, in which case only $\Delta_1^2>\Delta_0^2-\Delta_3^3$ is required.
\begin{figure}
\centerline{\includegraphics[width=.4\textwidth]{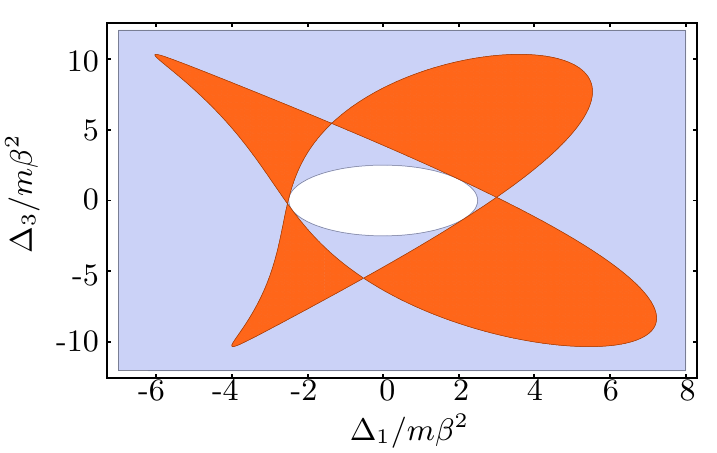}}
\caption{(color online) Topological phase diagram in the $\Delta_1-\Delta_3$ plane for $\Delta_0=2.5 m\beta^2, \alpha=2\beta, \gamma=0, V=3m\beta^2$,  and $\mu=t=10m\beta^2$. The light gray region fulfills the condition that $\det\Delta<0$, while the orange (dark gray) region corresponds the nontrivial topological phase.
\label{fig:Deltaz}}
\end{figure}
In Fig.~\ref{fig:phases} the structure of the transition is shown for the $\Delta_3=0$ case. The top panels show the spectrum before, at, and after a transition point. In the middle panels, the corresponding $Z_p$ trajectories in the complex plane are shown. Only when the parameter $V$ is between the two transition points does the trajectory encircle the origin, which is topologically different from the situation with $V<V_{c1}$ or $V>V_{c2}$. To illustrate the sign change of $Q$ the lower-left panel shows the real parts of $Z_{p1}$ and $Z_{p2}$, which have different signs only in the topologically nontrivial regime, in accordance with the criterion in Eq.~\eqref{Qdef}. Finally, the two lower-right panels show phase diagrams in two cuts of the parameter space, illustrating the robustness of the topological phase.

Now consider a different geometry with different intrawire pairings, i.e., $\Delta_3\neq0$. As an illustrative case, we can choose the parameters as $\Delta_0=\Delta_3=\Delta_1$, which means that the intrawire pairing in wire 2 is zero, while the interwire pairing is half of the intrawire pairing in wire 1. Further, taking $\alpha=\beta$ and $\gamma=0$, meaning that only wire 1 has spin-orbit coupling, the topological condition becomes $(4tV+\Delta_0^2)\Delta_0^2+4t^2V^2<8m\beta^2 t^2 (\mu+t+V)$. This could, for example, be a good approximation, if one wire is badly connected to the superconductor. The more general situation, when the intrawire pairing is finite in both wires, is shown in Fig.~\ref{fig:Deltaz}.

A key feature of the topological phase is the existence of localized states at the boundaries. In the following, we find the general form of these modes using an effective model containing the low-energy bands shown in Fig.~\ref{fig:geo} at the transition point. The general form of the effective 1D Hamiltonian follows by projection onto the low-energy bands (see Appendix):
\begin{equation}\label{Heff}
  H_{\mathrm{low}}= \left(\frac{p^2}{2m} -\tilde{\mu}\right)\tau_z + v(p\sigma_z-p_c)\tau_x
\end{equation}
where $p_c$ is the momentum at which the gap closes and $v$ and $\tilde{\mu}$ are effective parameters. This model describes a noncentrosymmetric superconductor because it contains both, $s$ and $p$-wave components of the superconducting pairing potential, and it is gapless when $\tilde{\mu}=\mu_c=p^2_c/2m$.
If we consider a hard boundary and that the wires to exist for $x>0$, it is easy to show from the secular equation that solutions exist for $\tau\sigma<0$ and $\tilde\mu>p_c^2/2m$. The two solutions then take the form
$\psi_{1(2)}(x)=\chi_{1(2)}f_{1(2)}(x)$,
in terms of the spinors $\chi_{1}=  (0,1,0,i)^T$, and $\chi_{2}=(1,0,-i,0)^T$, and with $f_{1}=f_{2}^*$ given by
\begin{equation}\label{fx}
  f_{1}(x)=A e^{-xmv} \sinh\left(x\sqrt{m^2v^2-2\tilde\mu m-2 i mv p_c }\right),
\end{equation}
where $A^2=8mv(\tilde\mu-\mu_c)/\sqrt{(2vp_c)^2+(mv^2-2\tilde\mu)^2}$.

The two zero modes $\psi_{1(2)}$ are \textit{not} Majorana bound states, because they are not eigenstates of $\mathcal{P}$, but only of $\mathcal{C}$. These solutions are the chiral symmetry-protected Jackiw-Rebbi-type topological solitons \cite{Jackiw1976,Su1979,Pershoguba2012}. We can, however, make linear combinations that are Majorana bound states. One example of a linear combinations that gives MBS (i.e. which fulfills $P\psi_M=\psi_M$) is
\begin{align}\label{MMODES}
  \psi_{M,1} =\frac{i\psi_{1}+\psi_{2}}{\sqrt{2}},\quad
  \psi_{M,2} =\frac{\psi_{1}+i\psi_{2}}{\sqrt{2}}.
\end{align}
These are MBS and transform to each other under TRS: $\mathcal{T} \psi_{M,2}= \psi_{M,1}$ and $\mathcal{T} \psi_{M,1}= -\psi_{M,2}$, which means that we have a Kramers pair of MBS.

Finally, we consider the effect of a Zeeman term that can split the two zero modes. The Hamiltonian \eqref{Heff} gets an additional time reversal-symmetry breaking term:
\begin{equation}\label{HZ}
  H_Z= \mathbf{B}\cdot \boldsymbol{\sigma}.
\end{equation}
If the field points along the spin-orbit direction the chiral symmetric states $\psi_1$ and $\psi_2$ are still eigenstates, but the degeneracy is lifted by 2$B_z$. A more interesting case is when the magnetic field is perpendicular to the spin-orbit direction, for example pointing in the $x$-direction. Figure \ref{figB} represents the topological phase diagram in this case. Three distinct phases correspond to different numbers, $N$, of MBS in each end of the effective 1D system. At zero magnetic field the nanowire belongs to the DIII topological symmetry class and at a finite magnetic field to the BDI class (with effective time-reversal symmetry $\mathcal{T} = \sigma_x\mathcal{K}$). Topological phase transitions to the phase with $N$ = 1 are associated with the gap closing at zero momentum and can be described by the equation $|B|$ = $\sqrt{(v p_c)^2+\tilde \mu^2}$. The transition between phases $N$ = 0 and $N$ = 2 is related to the gap closing at the Fermi momentum ($p$ = $\sqrt{2 m \tilde \mu}$) and can be described by the equation $v p_c$ = $\sqrt{B^2+2v ^2 m \tilde \mu }$. Note that disorder that breaks the effective TRS splits the $N=2$ MBSs, except at $B=0$ (which is the DIII situation studied above), while the $N=1$ regions are stable and merely reduce to class D.
\begin{figure}
\includegraphics[scale=1]{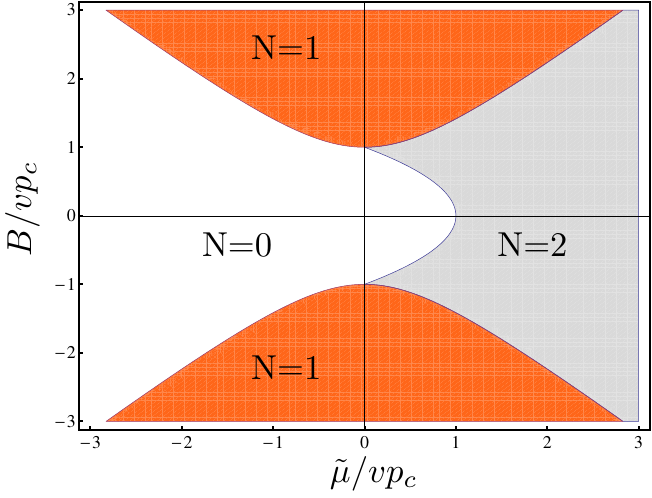}
\caption{Topological phase diagram for the low energy model (Eq.~\eqref{Heff}) in the presence of a magnetic field, assumed to be orthogonal to the $z$ axis, and for $p_{c}$ = $2vm$. Orange (dark gray) regions have single localized MBS, while the light gray region has MBS doublets.\label{figB}}
\end{figure}

To conclude, we have shown that a pair of time-reversal-symmetric nanowires proximity coupled to a superconductor can be driven into a nontrivial topological phase which supports a Kramers pair of Majorana bound states in each end. The key ingredients are interwire pairing and different spin-orbit interaction in the two wires. In the absence of interwire pairing, one needs intrawire pairing with different signs. With the assumption of parallel spin-orbit directions in the wires, the topological structure of the model could be determined from the AIII symmetric block diagonal parts of the full BdG Hamiltonian. However, we emphasize that the assumption of collinearity is not crucial for the existence of the topologically nontrivial phase. We have presented an analytical approach to find the topological invariant, which allows a general examination of the conditions for topological phases in systems using only ordinary $s$-wave
superconductors, proximity coupled to wires with spin-orbit coupling.

We thank Yuval Oreg for useful discussions. The Center for Quantum Devices is funded by the Danish National Research Foundation.

Note added in proof: A recent paper \cite{Haim2013} investigates the conditions for interaction-induced negative determinant of the pairing matrix in a two wire setup.

%


\begin{thebibliography}{37}%
\makeatletter
\providecommand \@ifxundefined [1]{%
 \@ifx{#1\undefined}
}%
\providecommand \@ifnum [1]{%
 \ifnum #1\expandafter \@firstoftwo
 \else \expandafter \@secondoftwo
 \fi
}%
\providecommand \@ifx [1]{%
 \ifx #1\expandafter \@firstoftwo
 \else \expandafter \@secondoftwo
 \fi
}%
\providecommand \natexlab [1]{#1}%
\providecommand \enquote  [1]{``#1''}%
\providecommand \bibnamefont  [1]{#1}%
\providecommand \bibfnamefont [1]{#1}%
\providecommand \citenamefont [1]{#1}%
\providecommand \href@noop [0]{\@secondoftwo}%
\providecommand \href [0]{\begingroup \@sanitize@url \@href}%
\providecommand \@href[1]{\@@startlink{#1}\@@href}%
\providecommand \@@href[1]{\endgroup#1\@@endlink}%
\providecommand \@sanitize@url [0]{\catcode `\\12\catcode `\$12\catcode
  `\&12\catcode `\#12\catcode `\^12\catcode `\_12\catcode `\%12\relax}%
\providecommand \@@startlink[1]{}%
\providecommand \@@endlink[0]{}%
\providecommand \url  [0]{\begingroup\@sanitize@url \@url }%
\providecommand \@url [1]{\endgroup\@href {#1}{\urlprefix }}%
\providecommand \urlprefix  [0]{URL }%
\providecommand \Eprint [0]{\href }%
\providecommand \doibase [0]{http://dx.doi.org/}%
\providecommand \selectlanguage [0]{\@gobble}%
\providecommand \bibinfo  [0]{\@secondoftwo}%
\providecommand \bibfield  [0]{\@secondoftwo}%
\providecommand \translation [1]{[#1]}%
\providecommand \BibitemOpen [0]{}%
\providecommand \bibitemStop [0]{}%
\providecommand \bibitemNoStop [0]{.\EOS\space}%
\providecommand \EOS [0]{\spacefactor3000\relax}%
\providecommand \BibitemShut  [1]{\csname bibitem#1\endcsname}%
\let\auto@bib@innerbib\@empty
\bibitem [{\citenamefont {Kitaev}(2001)}]{Kitaev2001}%
  \BibitemOpen
  \bibfield  {author} {\bibinfo {author} {\bibfnamefont {A.~Y.}\ \bibnamefont
  {Kitaev}},\ }\href {\doibase 10.1070/1063-7869/44/10S/S29} {\bibfield
  {journal} {\bibinfo  {journal} {Physics-Uspekhi}\ }\textbf {\bibinfo {volume}
  {44}},\ \bibinfo {pages} {131} (\bibinfo {year} {2001})}\BibitemShut
  {NoStop}%
\bibitem [{\citenamefont {Ivanov}(2001)}]{Ivanov2001}%
  \BibitemOpen
  \bibfield  {author} {\bibinfo {author} {\bibfnamefont {D.~A.}\ \bibnamefont
  {Ivanov}},\ }\href {\doibase 10.1103/PhysRevLett.86.268} {\bibfield
  {journal} {\bibinfo  {journal} {Phys.~Rev.~Lett.}\ }\textbf {\bibinfo
  {volume} {86}},\ \bibinfo {pages} {268} (\bibinfo {year} {2001})}\BibitemShut
  {NoStop}%
\bibitem [{\citenamefont {Nayak}\ \emph {et~al.}(2008)\citenamefont {Nayak},
  \citenamefont {Simon}, \citenamefont {Stern}, \citenamefont {Freedman},\ and\
  \citenamefont {{Das Sarma}}}]{Nayak2008}%
  \BibitemOpen
  \bibfield  {author} {\bibinfo {author} {\bibfnamefont {C.}~\bibnamefont
  {Nayak}}, \bibinfo {author} {\bibfnamefont {S.}~\bibnamefont {Simon}},
  \bibinfo {author} {\bibfnamefont {A.}~\bibnamefont {Stern}}, \bibinfo
  {author} {\bibfnamefont {M.}~\bibnamefont {Freedman}}, \ and\ \bibinfo
  {author} {\bibfnamefont {S.}~\bibnamefont {{Das Sarma}}},\ }\href {\doibase
  10.1103/RevModPhys.80.1083} {\bibfield  {journal} {\bibinfo  {journal}
  {Rev.~Mod.~Phys.}\ }\textbf {\bibinfo {volume} {80}},\ \bibinfo {pages}
  {1083} (\bibinfo {year} {2008})}\BibitemShut {NoStop}%
\bibitem [{\citenamefont {Alicea}(2012)}]{AliceaReview}%
  \BibitemOpen
  \bibfield  {author} {\bibinfo {author} {\bibfnamefont {J.}~\bibnamefont
  {Alicea}},\ }\href {\doibase 10.1088/0034-4885/75/7/076501} {\bibfield
  {journal} {\bibinfo  {journal} {Rep.~Prog.~Phys.}\ }\textbf {\bibinfo
  {volume} {75}},\ \bibinfo {pages} {076501} (\bibinfo {year}
  {2012})}\BibitemShut {NoStop}%
\bibitem [{Bee()}]{BeenakkerReview}%
  \BibitemOpen
  \href@noop {} {}\bibinfo {note} {C. W. J. Beenakker,
  arXiv:1112.1950}\BibitemShut {NoStop}%
\bibitem [{\citenamefont {Leijnse}\ and\ \citenamefont
  {Flensberg}(2012)}]{LeijnseReview}%
  \BibitemOpen
  \bibfield  {author} {\bibinfo {author} {\bibfnamefont {M.}~\bibnamefont
  {Leijnse}}\ and\ \bibinfo {author} {\bibfnamefont {K.}~\bibnamefont
  {Flensberg}},\ }\href {\doibase 10.1088/0268-1242/27/12/124003} {\bibfield
  {journal} {\bibinfo  {journal} {Semiconductor Science and Technology}\
  }\textbf {\bibinfo {volume} {27}},\ \bibinfo {pages} {124003} (\bibinfo
  {year} {2012})}\BibitemShut {NoStop}%
\bibitem [{\citenamefont {Hasan}\ and\ \citenamefont
  {Kane}(2010)}]{HasanKaneReview}%
  \BibitemOpen
  \bibfield  {author} {\bibinfo {author} {\bibfnamefont {M.~Z.}\ \bibnamefont
  {Hasan}}\ and\ \bibinfo {author} {\bibfnamefont {C.~L.}\ \bibnamefont
  {Kane}},\ }\href {\doibase 10.1103/RevModPhys.82.3045} {\bibfield  {journal}
  {\bibinfo  {journal} {Reviews of Modern Physics}\ }\textbf {\bibinfo {volume}
  {82}},\ \bibinfo {pages} {3045} (\bibinfo {year} {2010})}\BibitemShut
  {NoStop}%
\bibitem [{\citenamefont {Fu}\ and\ \citenamefont {Kane}(2008)}]{Fu2008}%
  \BibitemOpen
  \bibfield  {author} {\bibinfo {author} {\bibfnamefont {L.}~\bibnamefont
  {Fu}}\ and\ \bibinfo {author} {\bibfnamefont {C.~L.}\ \bibnamefont {Kane}},\
  }\href {\doibase 10.1103/PhysRevLett.100.096407} {\bibfield  {journal}
  {\bibinfo  {journal} {Phys.~Rev.~Lett.}\ }\textbf {\bibinfo {volume} {100}},\
  \bibinfo {pages} {096407} (\bibinfo {year} {2008})}\BibitemShut {NoStop}%
\bibitem [{\citenamefont {Lutchyn}\ \emph {et~al.}(2010)\citenamefont
  {Lutchyn}, \citenamefont {Sau},\ and\ \citenamefont {{Das
  Sarma}}}]{Lutchyn2010}%
  \BibitemOpen
  \bibfield  {author} {\bibinfo {author} {\bibfnamefont {R.~M.}\ \bibnamefont
  {Lutchyn}}, \bibinfo {author} {\bibfnamefont {J.~D.}\ \bibnamefont {Sau}}, \
  and\ \bibinfo {author} {\bibfnamefont {S.}~\bibnamefont {{Das Sarma}}},\
  }\href@noop {} {\bibfield  {journal} {\bibinfo  {journal} {Phys.~Rev.~Lett.}\
  }\textbf {\bibinfo {volume} {105}},\ \bibinfo {pages} {077001} (\bibinfo
  {year} {2010})}\BibitemShut {NoStop}%
\bibitem [{\citenamefont {Oreg}\ \emph {et~al.}(2010)\citenamefont {Oreg},
  \citenamefont {Refael},\ and\ \citenamefont {von Oppen}}]{Oreg2010}%
  \BibitemOpen
  \bibfield  {author} {\bibinfo {author} {\bibfnamefont {Y.}~\bibnamefont
  {Oreg}}, \bibinfo {author} {\bibfnamefont {G.}~\bibnamefont {Refael}}, \ and\
  \bibinfo {author} {\bibfnamefont {F.}~\bibnamefont {von Oppen}},\ }\href
  {\doibase 10.1103/PhysRevLett.105.177002} {\bibfield  {journal} {\bibinfo
  {journal} {Phys.~Rev.~Lett.}\ }\textbf {\bibinfo {volume} {105}},\ \bibinfo
  {pages} {177002} (\bibinfo {year} {2010})}\BibitemShut {NoStop}%
\bibitem [{\citenamefont {Potter}\ and\ \citenamefont
  {Lee}(2010)}]{Potter2010}%
  \BibitemOpen
  \bibfield  {author} {\bibinfo {author} {\bibfnamefont {A.~C.}\ \bibnamefont
  {Potter}}\ and\ \bibinfo {author} {\bibfnamefont {P.~A.}\ \bibnamefont
  {Lee}},\ }\href {\doibase 10.1103/PhysRevLett.105.227003} {\bibfield
  {journal} {\bibinfo  {journal} {Phys.~Rev.~Lett.}\ }\textbf {\bibinfo
  {volume} {105}},\ \bibinfo {pages} {227003} (\bibinfo {year} {2010})},\
  \Eprint {http://arxiv.org/abs/0010440} {arXiv:0010440 [arXiv:cond-mat]}
  \BibitemShut {NoStop}%
\bibitem [{\citenamefont {Lutchyn}\ \emph {et~al.}(2011)\citenamefont
  {Lutchyn}, \citenamefont {Stanescu},\ and\ \citenamefont {{Das
  Sarma}}}]{Lutchyn2011}%
  \BibitemOpen
  \bibfield  {author} {\bibinfo {author} {\bibfnamefont {R.~M.}\ \bibnamefont
  {Lutchyn}}, \bibinfo {author} {\bibfnamefont {T.~D.}\ \bibnamefont
  {Stanescu}}, \ and\ \bibinfo {author} {\bibfnamefont {S.}~\bibnamefont {{Das
  Sarma}}},\ }\href {\doibase 10.1103/PhysRevLett.106.127001} {\bibfield
  {journal} {\bibinfo  {journal} {Phys.~Rev.~Lett.}\ }\textbf {\bibinfo
  {volume} {106}},\ \bibinfo {pages} {127001} (\bibinfo {year}
  {2011})}\BibitemShut {NoStop}%
\bibitem [{\citenamefont {Mourik}\ \emph {et~al.}(2012)\citenamefont {Mourik},
  \citenamefont {Zuo}, \citenamefont {Frolov}, \citenamefont {Plissard},
  \citenamefont {Bakkers},\ and\ \citenamefont {Kouwenhoven}}]{Mourik2012}%
  \BibitemOpen
  \bibfield  {author} {\bibinfo {author} {\bibfnamefont {V.}~\bibnamefont
  {Mourik}}, \bibinfo {author} {\bibfnamefont {K.}~\bibnamefont {Zuo}},
  \bibinfo {author} {\bibfnamefont {S.~M.}\ \bibnamefont {Frolov}}, \bibinfo
  {author} {\bibfnamefont {S.~R.}\ \bibnamefont {Plissard}}, \bibinfo {author}
  {\bibfnamefont {E.~P. A.~M.}\ \bibnamefont {Bakkers}}, \ and\ \bibinfo
  {author} {\bibfnamefont {L.~P.}\ \bibnamefont {Kouwenhoven}},\ }\href
  {\doibase 10.1126/science.1222360} {\bibfield  {journal} {\bibinfo  {journal}
  {Science}\ }\textbf {\bibinfo {volume} {1003}},\ \bibinfo {pages} {1003}
  (\bibinfo {year} {2012})}\BibitemShut {NoStop}%
\bibitem [{\citenamefont {Das}\ \emph {et~al.}(2012)\citenamefont {Das},
  \citenamefont {Ronen}, \citenamefont {Most}, \citenamefont {Oreg},
  \citenamefont {Heiblum},\ and\ \citenamefont {Shtrikman}}]{Das2012}%
  \BibitemOpen
  \bibfield  {author} {\bibinfo {author} {\bibfnamefont {A.}~\bibnamefont
  {Das}}, \bibinfo {author} {\bibfnamefont {Y.}~\bibnamefont {Ronen}}, \bibinfo
  {author} {\bibfnamefont {Y.}~\bibnamefont {Most}}, \bibinfo {author}
  {\bibfnamefont {Y.}~\bibnamefont {Oreg}}, \bibinfo {author} {\bibfnamefont
  {M.}~\bibnamefont {Heiblum}}, \ and\ \bibinfo {author} {\bibfnamefont
  {H.}~\bibnamefont {Shtrikman}},\ }\href {\doibase 10.1038/nphys2479}
  {\bibfield  {journal} {\bibinfo  {journal} {Nature Physics}\ }\textbf
  {\bibinfo {volume} {8}},\ \bibinfo {pages} {887} (\bibinfo {year}
  {2012})}\BibitemShut {NoStop}%
\bibitem [{\citenamefont {Churchill}\ \emph {et~al.}(2013)\citenamefont
  {Churchill}, \citenamefont {Fatemi}, \citenamefont {Grove-Rasmussen},
  \citenamefont {Deng}, \citenamefont {Caroff}, \citenamefont {Xu},\ and\
  \citenamefont {Marcus}}]{Churchill2013}%
  \BibitemOpen
  \bibfield  {author} {\bibinfo {author} {\bibfnamefont {H.~O.~H.}\
  \bibnamefont {Churchill}}, \bibinfo {author} {\bibfnamefont {V.}~\bibnamefont
  {Fatemi}}, \bibinfo {author} {\bibfnamefont {K.}~\bibnamefont
  {Grove-Rasmussen}}, \bibinfo {author} {\bibfnamefont {M.~T.}\ \bibnamefont
  {Deng}}, \bibinfo {author} {\bibfnamefont {P.}~\bibnamefont {Caroff}},
  \bibinfo {author} {\bibfnamefont {H.~Q.}\ \bibnamefont {Xu}}, \ and\ \bibinfo
  {author} {\bibfnamefont {C.~M.}\ \bibnamefont {Marcus}},\ }\href {\doibase
  10.1103/PhysRevB.87.241401} {\bibfield  {journal} {\bibinfo  {journal}
  {Phys.~Rev.~B}\ }\textbf {\bibinfo {volume} {87}},\ \bibinfo {pages} {241401}
  (\bibinfo {year} {2013})}\BibitemShut {NoStop}%
\bibitem [{\citenamefont {Alicea}\ \emph {et~al.}(2011)\citenamefont {Alicea},
  \citenamefont {Oreg}, \citenamefont {Refael}, \citenamefont {von Oppen},\
  and\ \citenamefont {Fisher}}]{Alicea2011}%
  \BibitemOpen
  \bibfield  {author} {\bibinfo {author} {\bibfnamefont {J.}~\bibnamefont
  {Alicea}}, \bibinfo {author} {\bibfnamefont {Y.}~\bibnamefont {Oreg}},
  \bibinfo {author} {\bibfnamefont {G.}~\bibnamefont {Refael}}, \bibinfo
  {author} {\bibfnamefont {F.}~\bibnamefont {von Oppen}}, \ and\ \bibinfo
  {author} {\bibfnamefont {M.~P.~A.}\ \bibnamefont {Fisher}},\ }\href {\doibase
  10.1038/nphys1915} {\bibfield  {journal} {\bibinfo  {journal} {Nature
  Physics}\ }\textbf {\bibinfo {volume} {7}},\ \bibinfo {pages} {412} (\bibinfo
  {year} {2011})}\BibitemShut {NoStop}%
\bibitem [{\citenamefont {Halperin}\ \emph {et~al.}(2012)\citenamefont
  {Halperin}, \citenamefont {Oreg}, \citenamefont {Stern}, \citenamefont
  {Refael}, \citenamefont {Alicea},\ and\ \citenamefont {von
  Oppen}}]{Halperin2012}%
  \BibitemOpen
  \bibfield  {author} {\bibinfo {author} {\bibfnamefont {B.~I.}\ \bibnamefont
  {Halperin}}, \bibinfo {author} {\bibfnamefont {Y.}~\bibnamefont {Oreg}},
  \bibinfo {author} {\bibfnamefont {A.}~\bibnamefont {Stern}}, \bibinfo
  {author} {\bibfnamefont {G.}~\bibnamefont {Refael}}, \bibinfo {author}
  {\bibfnamefont {J.}~\bibnamefont {Alicea}}, \ and\ \bibinfo {author}
  {\bibfnamefont {F.}~\bibnamefont {von Oppen}},\ }\href {\doibase
  10.1103/PhysRevB.85.144501} {\bibfield  {journal} {\bibinfo  {journal}
  {Phys.~Rev.~B}\ }\textbf {\bibinfo {volume} {85}},\ \bibinfo {pages} {144501}
  (\bibinfo {year} {2012})}\BibitemShut {NoStop}%
\bibitem [{\citenamefont {Fulga}\ \emph {et~al.}(2011)\citenamefont {Fulga},
  \citenamefont {Hassler}, \citenamefont {Akhmerov},\ and\ \citenamefont
  {Beenakker}}]{Fulga2011}%
  \BibitemOpen
  \bibfield  {author} {\bibinfo {author} {\bibfnamefont {I.~C.}\ \bibnamefont
  {Fulga}}, \bibinfo {author} {\bibfnamefont {F.}~\bibnamefont {Hassler}},
  \bibinfo {author} {\bibfnamefont {A.~R.}\ \bibnamefont {Akhmerov}}, \ and\
  \bibinfo {author} {\bibfnamefont {C.~W.~J.}\ \bibnamefont {Beenakker}},\
  }\href {\doibase 10.1103/PhysRevB.83.155429} {\bibfield  {journal} {\bibinfo
  {journal} {Physical Review B}\ }\textbf {\bibinfo {volume} {83}},\ \bibinfo
  {pages} {155429} (\bibinfo {year} {2011})}\BibitemShut {NoStop}%
\bibitem [{\citenamefont {Stanescu}\ \emph {et~al.}(2011)\citenamefont
  {Stanescu}, \citenamefont {Lutchyn},\ and\ \citenamefont {{Das
  Sarma}}}]{Stanescu2011}%
  \BibitemOpen
  \bibfield  {author} {\bibinfo {author} {\bibfnamefont {T.~D.}\ \bibnamefont
  {Stanescu}}, \bibinfo {author} {\bibfnamefont {R.~M.}\ \bibnamefont
  {Lutchyn}}, \ and\ \bibinfo {author} {\bibfnamefont {S.}~\bibnamefont {{Das
  Sarma}}},\ }\href {\doibase 10.1103/PhysRevB.84.144522} {\bibfield  {journal}
  {\bibinfo  {journal} {Physical Review B}\ }\textbf {\bibinfo {volume} {84}},\
  \bibinfo {pages} {144522} (\bibinfo {year} {2011})}\BibitemShut {NoStop}%
\bibitem [{\citenamefont {Wong}\ and\ \citenamefont {Law}(2012)}]{Wong2012}%
  \BibitemOpen
  \bibfield  {author} {\bibinfo {author} {\bibfnamefont {C.~L.~M.}\
  \bibnamefont {Wong}}\ and\ \bibinfo {author} {\bibfnamefont {K.~T.}\
  \bibnamefont {Law}},\ }\href {\doibase 10.1103/PhysRevB.86.184516} {\bibfield
   {journal} {\bibinfo  {journal} {Phys.~Rev.~B}\ }\textbf {\bibinfo {volume}
  {86}},\ \bibinfo {pages} {184516} (\bibinfo {year} {2012})}\BibitemShut
  {NoStop}%
\bibitem [{\citenamefont {Zhang}\ \emph {et~al.}(2013)\citenamefont {Zhang},
  \citenamefont {Kane},\ and\ \citenamefont {Mele}}]{Zhang2013}%
  \BibitemOpen
  \bibfield  {author} {\bibinfo {author} {\bibfnamefont {F.}~\bibnamefont
  {Zhang}}, \bibinfo {author} {\bibfnamefont {C.~L.}\ \bibnamefont {Kane}}, \
  and\ \bibinfo {author} {\bibfnamefont {E.~J.}\ \bibnamefont {Mele}},\ }\href
  {\doibase 10.1103/PhysRevLett.111.056402} {\bibfield  {journal} {\bibinfo
  {journal} {Phys.~Rev.~Lett.}\ }\textbf {\bibinfo {volume} {111}},\ \bibinfo
  {pages} {056402} (\bibinfo {year} {2013})}\BibitemShut {NoStop}%
\bibitem [{\citenamefont {Nakosai}\ \emph {et~al.}(2013)\citenamefont
  {Nakosai}, \citenamefont {Budich}, \citenamefont {Tanaka}, \citenamefont
  {Trauzettel},\ and\ \citenamefont {Nagaosa}}]{Nakosai2013}%
  \BibitemOpen
  \bibfield  {author} {\bibinfo {author} {\bibfnamefont {S.}~\bibnamefont
  {Nakosai}}, \bibinfo {author} {\bibfnamefont {J.~C.}\ \bibnamefont {Budich}},
  \bibinfo {author} {\bibfnamefont {Y.}~\bibnamefont {Tanaka}}, \bibinfo
  {author} {\bibfnamefont {B.}~\bibnamefont {Trauzettel}}, \ and\ \bibinfo
  {author} {\bibfnamefont {N.}~\bibnamefont {Nagaosa}},\ }\href@noop {}
  {\bibfield  {journal} {\bibinfo  {journal} {Phys.~Rev.~Lett.}\ }\textbf
  {\bibinfo {volume} {110}},\ \bibinfo {pages} {117002} (\bibinfo {year}
  {2013})}\BibitemShut {NoStop}%
\bibitem [{\citenamefont {Nakosai}\ \emph {et~al.}(2012)\citenamefont
  {Nakosai}, \citenamefont {Tanaka},\ and\ \citenamefont
  {Nagaosa}}]{Nakosai2012}%
  \BibitemOpen
  \bibfield  {author} {\bibinfo {author} {\bibfnamefont {S.}~\bibnamefont
  {Nakosai}}, \bibinfo {author} {\bibfnamefont {Y.}~\bibnamefont {Tanaka}}, \
  and\ \bibinfo {author} {\bibfnamefont {N.}~\bibnamefont {Nagaosa}},\ }\href
  {\doibase 10.1103/PhysRevLett.108.147003} {\bibfield  {journal} {\bibinfo
  {journal} {Phys.~Rev.~Lett.}\ }\textbf {\bibinfo {volume} {108}},\ \bibinfo
  {pages} {147003} (\bibinfo {year} {2012})}\BibitemShut {NoStop}%
\bibitem [{\citenamefont {Deng}\ \emph {et~al.}(2012)\citenamefont {Deng},
  \citenamefont {Viola},\ and\ \citenamefont {Ortiz}}]{Deng2012}%
  \BibitemOpen
  \bibfield  {author} {\bibinfo {author} {\bibfnamefont {S.}~\bibnamefont
  {Deng}}, \bibinfo {author} {\bibfnamefont {L.}~\bibnamefont {Viola}}, \ and\
  \bibinfo {author} {\bibfnamefont {G.}~\bibnamefont {Ortiz}},\ }\href@noop {}
  {\bibfield  {journal} {\bibinfo  {journal} {Phys. \ Rev. \ Lett.}\ }\textbf
  {\bibinfo {volume} {108}},\ \bibinfo {pages} {036803} (\bibinfo {year}
  {2012})}\BibitemShut {NoStop}%
\bibitem [{\citenamefont {Keselman}\ \emph {et~al.}(2013)\citenamefont
  {Keselman}, \citenamefont {Fu}, \citenamefont {Stern},\ and\ \citenamefont
  {Berg}}]{Keselman2013}%
  \BibitemOpen
  \bibfield  {author} {\bibinfo {author} {\bibfnamefont {A.}~\bibnamefont
  {Keselman}}, \bibinfo {author} {\bibfnamefont {L.}~\bibnamefont {Fu}},
  \bibinfo {author} {\bibfnamefont {A.}~\bibnamefont {Stern}}, \ and\ \bibinfo
  {author} {\bibfnamefont {E.}~\bibnamefont {Berg}},\ }\href {\doibase
  10.1103/PhysRevLett.111.116402} {\bibfield  {journal} {\bibinfo  {journal}
  {Phys.~Rev.~Lett.}\ }\textbf {\bibinfo {volume} {111}},\ \bibinfo {pages}
  {116402} (\bibinfo {year} {2013})}\BibitemShut {NoStop}%
\bibitem [{Liu()}]{Liu2013}%
  \BibitemOpen
  \href@noop {} {}\bibinfo {note} {X.-J. Liu, C. L. M. Wong, and K. T. Law,
  arXiv:1304.3765}\BibitemShut {NoStop}%
\bibitem [{Chu()}]{Chung2012}%
  \BibitemOpen
  \href@noop {} {}\bibinfo {note} {S. B. Chung, J. Horowitz, and X.-L. Qi,
  arXiv:1208.3928}\BibitemShut {NoStop}%
\bibitem [{\citenamefont {Gangadharaiah}\ \emph {et~al.}(2011)\citenamefont
  {Gangadharaiah}, \citenamefont {Braunecker}, \citenamefont {Simon},\ and\
  \citenamefont {Loss}}]{Gangadharaiah2011}%
  \BibitemOpen
  \bibfield  {author} {\bibinfo {author} {\bibfnamefont {S.}~\bibnamefont
  {Gangadharaiah}}, \bibinfo {author} {\bibfnamefont {B.}~\bibnamefont
  {Braunecker}}, \bibinfo {author} {\bibfnamefont {P.}~\bibnamefont {Simon}}, \
  and\ \bibinfo {author} {\bibfnamefont {D.}~\bibnamefont {Loss}},\ }\href
  {\doibase 10.1103/PhysRevLett.107.036801} {\bibfield  {journal} {\bibinfo
  {journal} {Phys.~Rev.~Lett.}\ }\textbf {\bibinfo {volume} {107}},\ \bibinfo
  {pages} {036801} (\bibinfo {year} {2011})}\BibitemShut {NoStop}%
\bibitem [{\citenamefont {Stoudenmire}\ \emph {et~al.}(2011)\citenamefont
  {Stoudenmire}, \citenamefont {Alicea}, \citenamefont {Starykh},\ and\
  \citenamefont {Fisher}}]{Stoudenmire2011}%
  \BibitemOpen
  \bibfield  {author} {\bibinfo {author} {\bibfnamefont {E.~M.}\ \bibnamefont
  {Stoudenmire}}, \bibinfo {author} {\bibfnamefont {J.}~\bibnamefont {Alicea}},
  \bibinfo {author} {\bibfnamefont {O.~A.}\ \bibnamefont {Starykh}}, \ and\
  \bibinfo {author} {\bibfnamefont {M.~P.~A.}\ \bibnamefont {Fisher}},\ }\href
  {\doibase 10.1103/PhysRevB.84.014503} {\bibfield  {journal} {\bibinfo
  {journal} {Phys.~Rev.~B}\ }\textbf {\bibinfo {volume} {84}},\ \bibinfo
  {pages} {014503} (\bibinfo {year} {2011})}\BibitemShut {NoStop}%
\bibitem [{\citenamefont {Recher}\ and\ \citenamefont
  {Loss}(2002)}]{Recher2002}%
  \BibitemOpen
  \bibfield  {author} {\bibinfo {author} {\bibfnamefont {P.}~\bibnamefont
  {Recher}}\ and\ \bibinfo {author} {\bibfnamefont {D.}~\bibnamefont {Loss}},\
  }\href {\doibase 10.1103/PhysRevB.65.165327} {\bibfield  {journal} {\bibinfo
  {journal} {Phys.~Rev.~B}\ }\textbf {\bibinfo {volume} {65}},\ \bibinfo
  {pages} {165327} (\bibinfo {year} {2002})}\BibitemShut {NoStop}%
\bibitem [{sym()}]{sym}%
  \BibitemOpen
  \href@noop {} {}\bibinfo {note} {In class DIII, we have
  $\mathcal{T}\mathcal{H}(p)\mathcal{T}^{\dagger} = \mathcal{H}(-p),
  \mathcal{P}\mathcal{H}(p)\mathcal{P}^{\dagger} = -\mathcal{H}(-p)$, and $
  \mathcal{C}\mathcal{H}(p)\mathcal{C}^{\dagger} =
  -\mathcal{H}(p)$.}\BibitemShut {Stop}%
\bibitem [{kes()}]{keselmannote}%
  \BibitemOpen
  \href@noop {} {}\bibinfo {note} {For example for the model in
  Refs.~\onlinecite{Keselman2013,Deng2012}, which is equivalent to ours if
  $\Delta_1=\alpha=V=\Delta_0=0$, the double degeneracy of the bands makes it
  possible to off-diagonalize the Hamiltonian into $\mathcal{P}$-symmetric
  blocks and thus use the Kitaev Pfaffian criterion.}\BibitemShut {Stop}%
\bibitem [{\citenamefont {Tewari}\ and\ \citenamefont
  {Sau}(2012)}]{Tewari2012}%
  \BibitemOpen
  \bibfield  {author} {\bibinfo {author} {\bibfnamefont {S.}~\bibnamefont
  {Tewari}}\ and\ \bibinfo {author} {\bibfnamefont {J.~D.}\ \bibnamefont
  {Sau}},\ }\href {\doibase 10.1103/PhysRevLett.109.150408} {\bibfield
  {journal} {\bibinfo  {journal} {Phys.~Rev.~Lett.}\ }\textbf {\bibinfo
  {volume} {109}},\ \bibinfo {pages} {150408} (\bibinfo {year}
  {2012})}\BibitemShut {NoStop}%
\bibitem [{\citenamefont {Jackiw}\ and\ \citenamefont
  {Rebbi}(1976)}]{Jackiw1976}%
  \BibitemOpen
  \bibfield  {author} {\bibinfo {author} {\bibfnamefont {R.}~\bibnamefont
  {Jackiw}}\ and\ \bibinfo {author} {\bibfnamefont {C.}~\bibnamefont {Rebbi}},\
  }\href {\doibase 10.1103/PhysRevD.13.3398} {\bibfield  {journal} {\bibinfo
  {journal} {Phys. Rev. D}\ }\textbf {\bibinfo {volume} {13}},\ \bibinfo
  {pages} {3398} (\bibinfo {year} {1976})}\BibitemShut {NoStop}%
\bibitem [{\citenamefont {Su}\ \emph {et~al.}(1979)\citenamefont {Su},
  \citenamefont {Schrieffer},\ and\ \citenamefont {Heeger}}]{Su1979}%
  \BibitemOpen
  \bibfield  {author} {\bibinfo {author} {\bibfnamefont {W.~P.}\ \bibnamefont
  {Su}}, \bibinfo {author} {\bibfnamefont {J.~R.}\ \bibnamefont {Schrieffer}},
  \ and\ \bibinfo {author} {\bibfnamefont {A.~J.}\ \bibnamefont {Heeger}},\
  }\href {\doibase 10.1103/PhysRevLett.42.1698} {\bibfield  {journal} {\bibinfo
   {journal} {Phys.~Rev.~Lett.}\ }\textbf {\bibinfo {volume} {42}},\ \bibinfo
  {pages} {1698} (\bibinfo {year} {1979})}\BibitemShut {NoStop}%
\bibitem [{\citenamefont {Pershoguba}\ and\ \citenamefont
  {Yakovenko}(2012)}]{Pershoguba2012}%
  \BibitemOpen
  \bibfield  {author} {\bibinfo {author} {\bibfnamefont {S.~S.}\ \bibnamefont
  {Pershoguba}}\ and\ \bibinfo {author} {\bibfnamefont {V.~M.}\ \bibnamefont
  {Yakovenko}},\ }\href {\doibase 10.1103/PhysRevB.86.075304} {\bibfield
  {journal} {\bibinfo  {journal} {Phys.~Rev.~B}\ }\textbf {\bibinfo {volume}
  {86}},\ \bibinfo {pages} {075304} (\bibinfo {year} {2012})}\BibitemShut
  {NoStop}%
\bibitem [{Hai()}]{Haim2013}%
  \BibitemOpen
  \href@noop {} {}\bibinfo {note} {A. Haim, A. Keselman, E. Berg, Y. Oreg,
  arXiv:1310.4525}\BibitemShut {NoStop}%
\end{thebibliography}

\appendix*

\section{Appendix}

\subsection{Derivation of the effective low energy model}

In this section, we derive the effective low-energy model for the lowest bands shown in Fig. 1 in the main text. To keep the derivation relatively simple, we consider a reduced version with $\Delta_3=\alpha=\gamma=0$, which however does not change the form of the final low-energy model. We thus consider the two-channel model:
\begin{align}
H_{1}&=\left( \frac{p^{2}}{2m}-\mu +V\lambda _{z}+t\lambda _{x}+\beta p\sigma _{z}\lambda _{z}\right) \tau _{z}\notag\\
&\quad +\left( \Delta _{0}+\Delta_{1}\lambda _{x}\right) \tau _{x}.  \label{H1}
\end{align}
Because the pairing term does not commute with the first electron-hole part this is in general an 8x8 matrix. To make analytical progress possible, we will assume that the asymmetry and the spin-orbit couplings are weak so that we can treat the terms containing $\lambda_z$ as a perturbation:
\begin{equation}
H_{1}^{\prime }=\left( V+\beta p\sigma _{z}\right) \lambda _{z}\tau _{z},
\label{Hpert}
\end{equation}
which means that the unperturbed part of the Hamiltonian is now diagonal in eigenstates
of $\lambda _{x}$. With $P_{\pm }=(1\pm \lambda _{x})/2$ being projection
operators to the eigenstates of $\lambda _{x}$ with eigenvalues $\pm 1,$ the
zeroth order low energy Hamiltonian is thus
\begin{equation}
H_{1,\mathrm{low}}^{(0)}=\left\{ \left( \frac{p^{2}}{2m}-\mu -t\right) \tau
_{z}+\left( \Delta _{0}-\Delta _{1}\right) \tau _{x}\right\} P_{-},
\label{H1low}
\end{equation}
while the high energy part of the unperturbed Hamiltonian is
\begin{equation}
H_{1,\mathrm{high}}^{(0)}=\left\{ \left( \frac{p^{2}}{2m}-\mu +t\right) \tau
_{z}+(\Delta _{0}+\Delta _{1})\tau _{x}\right\} P_{+},  \label{H1high}
\end{equation}%
Second order perturbation theory now gives a correction to the low-energy part
\begin{equation}
H_{1}^{(2)}=P_{-}H_{1}^{\prime }\left[ E_{0}-H_{1,\mathrm{high}}^{(0)}\right]
^{-1}H_{1}^{\prime }P_{-},
\end{equation}%
where $E_{0}$ is the energy of the unperturbed low-energy state. Below, we will see that
Fermi point is not renormalized up to linear order in the perturbation, the
condition for a gap closing at the transition to a topological
superconductor happens near $p=\sqrt{2m(\mu +t)}$ and, therefore, we can
neglect $E_{0}$, assuming that $\Delta _{0}-\Delta _{1}\ll \sqrt{%
4t^{2}+(\Delta _{0}+\Delta _{1})^{2}}.$ Furthermore, since $\lambda _{z}$
and $\lambda _{y}$ both flip between the eigenstates of $\lambda _{x}$ ($%
\lambda _{z}P_{\mp }=P_{\pm }$) the second order correction can be written as%
\begin{align}
H_{1}^{(2)}&\approx -P_{-}\left( V+\beta p\sigma _{z}\right) \tau _{z}\left[
2t\tau _{z}+(\Delta _{0}+\Delta _{1})\tau _{x}\right] ^{-1}\notag\\
& \quad\times\tau _{z}\left(V+\beta p\sigma _{z}\right) P_{-}.
\end{align}
Combining this with the unperturbed low energy Hamiltonian, we get the final effective
low-energy Hamiltonian:
\begin{equation}
H_{1,\mathrm{low}}\approx \left( \frac{p^{2}}{2m}-\mu -t-\delta (p)\right)
\tau _{z}+\left( \Delta_s +p\sigma _{z}\Delta _{p}\right) \tau _{x},
\label{H1lowfinal}
\end{equation}%
where
\begin{align}
\Delta_s &=\Delta _{0}-\Delta _{1}+\frac{V^{2}+p^{2}\beta ^{2}}{D},\\
\Delta _{p}&=2\frac{\beta V}{D},\\
D &=\frac{4t^{2}+\left(
\Delta _{0}+\Delta _{1}\right) ^{2}}{\left( \Delta _{0}+\Delta _{1}\right) },\\
\delta (p)&=\frac{2t\left( V+\beta p\sigma _{z}\right) ^{2}}{4t^{2}+\left( \Delta _{0}+\Delta _{1}\right) ^{2}}.  \label{DD}
\end{align}
We have thus mapped the model to two decoupled 1D models, one for spin up and one for spin down. The two models are related by TRS and map to each other by $\sigma _{z}\rightarrow -\sigma _{z}$ and $p\rightarrow -p$. They can undergo a transition from a trivial to a topological p-wave superconductor, when the gap changes sign, which happens when
\begin{equation}
\Delta_s \pm p_{F}\Delta _{p}=0,  \label{gapless}
\end{equation}%
where $p_{F}$ is determined by $p_{F}=\sqrt{2m\left( \mu +t+\delta
(p_{F}\right) }.$

To put this condition in the context of a topological quantum number for a p-wave superconductor, we transform the Hamiltonian in (\ref{H1lowfinal}) by shifting $p$ as $p=k-\sigma _{z}\Delta_s /\Delta _{p},$which leads to%
\begin{align}
H_{1,\mathrm{low}}&=\left( \frac{k^{2}}{2m}-\mu -t-\delta +\frac{\Delta_s^{2}}{%
2m\Delta_{p}^{2}}-k\sigma _{z}\frac{\Delta_s }{m\Delta _{p}}\right) \tau_{z}\notag\\
\quad&+k\Delta _{p}\sigma _{z}\tau _{x}.
\end{align}
The condition for the 1D p-wave superconductor to be in the topological phase is that the total chemical potential is positive and hence
\begin{equation}
\mu +t+\delta -\frac{\Delta_s^{2}}{2m\Delta _{p}^{2}}>0.
\end{equation}%
The transition point thus agrees with the condition in Eq. (\ref{gapless})
for the gap to close.

For small $V_{0}+\beta p\sigma _{z}$, we neglect $\delta $ and get the condition%
\begin{equation}
2m\Delta _{p}^{2}\left( \mu +t\right) -\Delta_s ^{2}>0,
\end{equation}%
or
\begin{equation}
8m\left( \mu +t\right) \beta ^{2}V^{2}>\left( \left( \Delta _{0}-\Delta
_{1}\right) D+V^{2}+2m\left( \mu +t\right) \beta ^{2}\right) ^{2}.
\end{equation}%
For $t=0$, this becomes
\begin{equation}
K_{-}<\Delta _{1}<K_{+},
\end{equation}%
with
\begin{equation}
K_{\pm }=\sqrt{\Delta _{0}^{2}+\left( V\pm \sqrt{2m\mu }\beta \right) ^{2}}.
\end{equation}
We see that the condition for $\Delta _{1}$ always requires $\Delta _{1}>\Delta _{0}.$

\subsection{Determinant of pairing matrix with coupling to a conventional superconductor}

Here we show that the pairing matrix for a non-interacting system coupled to a conventional $s$-wave superconductor with a positive order parameter has a positive determinant. 

For a normal system coupled to a superconductor one can integrate out the superconducting electrons, which to second order in tunnel coupling and for energies much smaller than the gap gives an effective pairing
\begin{equation}\label{Deltaxx}
  \Delta_{ij}=t_it_j\sum_\alpha\varphi_\alpha(x_i)\varphi_\alpha(x_j)\frac{\Delta_\alpha}{2E_\alpha},
\end{equation}
where $t_i$ is the tunnel coupling to channel $i$ (assumed energy independent), $\varphi_\alpha(x_i)$ is the electron wave function at the position of the wire $i$. Defining $\chi_\alpha(i)= t_i\varphi_\alpha(x_i) \sqrt{\Delta_\alpha/2E_\alpha}$, we can write this as
\begin{equation}\label{Deltaxx}
  \Delta_{ij}=\sum_\alpha \chi_\alpha(i) \chi_\alpha(j),
\end{equation}
from which it follows that $\Delta_{11}\Delta_{22}>\Delta_{12}\Delta_{21}$. With interactions this proof is no longer valid, because interactions, even at a mean-field level, renormalize the diagonal and off-diagonal parts differently.

\subsection{Condition on pairing matrix for a topological superconductor}

In this part we show that the determinant of the pairing matrix must be negative in order for the system to be topologically nontrivial.  Starting with our topological invariant $Z_p=\det(\Delta+i\mathcal{H}^0_{p,\uparrow})$, one can rotate the matrix such that $\Delta$ is diagonal, with eigenvalues $a$ and $b$. After rotation, $Z_p$ has the form
\begin{equation}\label{Zpform}
  Z_p=det \left(
            \begin{array}{cc}
              a+ix & iz \\
              iz & b+iy \\
            \end{array}
          \right)= ab+z^2-xy +i (ay+bx).
\end{equation}
We can then find the crossings with the real axis by setting $y=-bx/a$ (which gives two solutions for $p=p_{1(2)}$, as explained in the main text), so that
\begin{equation}\label{ReZp}
  Z_{p_1(2)}= ab+z^2+x^2b/a.
\end{equation}
Now it is clear that a necessary condition for this to be negative is $ab<0$, or equivalently $\det \Delta<0$.

\end{document}